\documentclass[sigconf]{acmart}

\setcopyright{none}

\renewcommand\footnotetextcopyrightpermission[1]{} 
\AtBeginDocument{%
  }

\setcopyright{acmlicensed}
\copyrightyear{2018}
\acmYear{2018}
\acmDOI{XXXXXXX.XXXXXXX}
\acmConference[Conference acronym 'XX]{Make sure to enter the correct
  conference title from your rights confirmation email}{June 03--05,
  2018}{Woodstock, NY}
\acmISBN{978-1-4503-XXXX-X/2018/06}

\usepackage{rotating}   
\usepackage{multirow}
\usepackage{subcaption}
\usepackage{enumitem}
\begin{document}

\title{ButterMamba: Butterworth-Enhanced Spatial-Temporal Mamba
for Efficient Traffic Flow Prediction}



\author{Limiao Zhang}
\affiliation{%
  \institution{State Key Laboratory of Opto‑Electronic Information Acquisition and Protection Technology, Anhui University}
  \city{Hefei}
  \state{Anhui}
  \postcode{230601}
  \country{China}
}
\email{zhanglm@ahu.edu.cn}

\author{Yuhui Lu}
\affiliation{%
  \institution{Institute of Physical Science and Information Technology, Anhui University}
  \city{Hefei}
  \state{Anhui}
  \postcode{230601}
  \country{China}
}
\email{q23301225@stu.ahu.edu.cn}

\author{Jie Gao}
\affiliation{%
  \institution{School of Innovation and Entrepreneurship, Shandong University}
  \city{Qingdao}
  \state{Shandong}
  \postcode{266237}
  \country{China}
}
\email{littlehope@sdu.edu.cn}

\author{Hao Jiang}
\affiliation{%
  \institution{School of Computer Science and Technology, Anhui University}
  \city{Hefei}
  \state{Anhui}
  \postcode{230601}
  \country{China}
}
\email{haojiang@ahu.edu.cn}

 \author{Haiping Ma}
\affiliation{%
  \institution{Institute of Physical Science and Information Technology, Anhui University}
  \city{Hefei}
  \state{Anhui}
  \postcode{230601}
  \country{China}
}
\email{hpma@ahu.edu.cn}

\author{Xingyi Zhang*}
\affiliation{%
  \institution{School of Computer Science and Technology, Anhui University}
  \city{Hefei}
  \state{Anhui}
  \postcode{230601}
  \country{China}
}
\email{xyzhanghust@gmail.com}

\begin{abstract}
Accurate traffic flow prediction is fundamental to intelligent transportation systems, playing a pivotal role in urban mobility optimization and smart city development. While Graph Neural Networks (GNNs) integrated with time series forecasting have emerged as promising solutions, two critical limitations persist: (1) the quadratic complexity of attention-based architectures hinders real-time deployment in large-scale networks, and (2) high-frequency noise in sensor data significantly degrades prediction reliability. These challenges are particularly acute in metropolitan scenarios where both computational efficiency and noise robustness are paramount. To address these limitations, we introduce \textbf{ButterMamba}, a novel and efficient framework based on State Space Models (SSMs). ButterMamba consists of two key components: (1) a Butterworth Spectral Filtering module that preprocesses the data by removing high-frequency noise, allowing the model to focus on significant underlying trends, and (2) a Spatial-Temporal State Mixer that uses a parallel Mamba architecture to efficiently capture both long-range temporal dependencies and complex spatial correlations across the road network. By decoupling noise filtering from spatial-temporal modeling, ButterMamba achieves superior predictive accuracy with linear computational complexity. Extensive experiments on three public datasets demonstrate that ButterMamba not only outperforms existing state-of-the-art models in terms of prediction accuracy but also considerably reduces training time and memory usage.

\end{abstract}




\maketitle

\section{Introduction}
As part of Intelligent Transportation Systems (ITS) \cite{Yin2015}, urban traffic flow prediction has gained increasing attention with the rapid development of intelligent transportation systems \cite{6894591}. For example, predicting future traffic flow on highways can help transportation authorities implement dynamic speed limits to prevent stop-and-go waves. 

The research on traffic flow prediction has witnessed a significant technical evolution, moving from statistical models to sophisticated deep learning architectures. Early approaches relied on classical time-series models like ARIMA and Kalman filtering  \cite{shekhar2007adaptive, moreira2013predicting, lippi2013short}. While valuable, these models were limited in their ability to capture the complex non-linearities and, more importantly, the spatial dependencies inherent in road networks. The first wave of deep learning introduced Recurrent Neural Networks (RNNs) and their variants, such as LSTM and GRU \cite{li2020knowledge,luo2018multivariate}, to model temporal dynamics, alongside Convolutional Neural Networks (CNNs) \cite{Zhang_Zheng_Qi_2017} to capture spatial correlations by treating road networks as grid-like structures. With advancements in Graph Neural Networks (GNNs), Spatio-Temporal Graph Neural Networks (STGNNs) emerged as a paradigm to effectively capture non-Euclidean spatial dependencies \cite{graphwang}. Representative frameworks include the Spatio-Temporal Graph Convolutional Network (STGCN) \cite{STGCN}, the Diffusion Convolutional Recurrent Neural Network (DCRNN) \cite{li2017diffusion}, Graph WaveNet \cite{wu2019graph} and so on. 

Recently, Transformer-based architectures have achieved excellent performance. Models like PDFormer \cite{PDFormer} leverage self-attention mechanisms to capture dynamic, long-range dependencies. 
However, the computational complexity of the self-attention mechanism scales quadratically with the input sequence length, becoming a significant bottleneck for long-term forecasting and large-scale networks. While numerous studies have proposed more efficient Transformer variants \cite{fasttransformer}, these optimizations often come at the cost of reduced predictive accuracy. This highlights a persistent trade-off between model efficiency and performance, leaving a critical gap for a solution that is both fast and highly accurate.

\begin{figure}[h]
  \centering
  \begin{minipage}[b]{0.45\textwidth}
    \centering
    \includegraphics[width=1.0\linewidth]{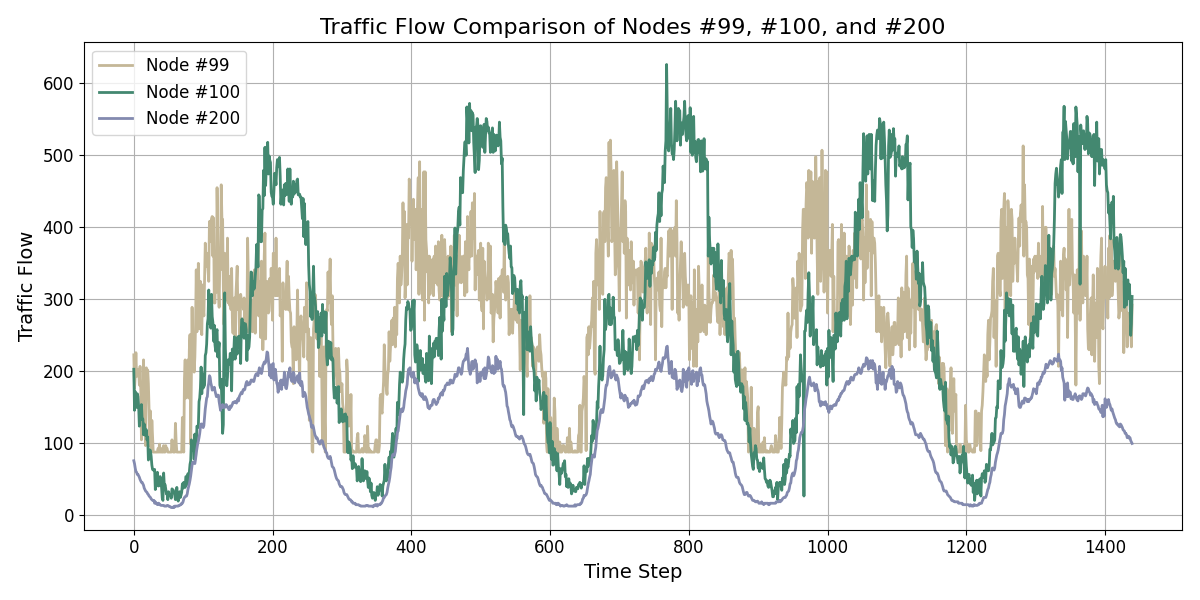}
  \end{minipage}
  
  \caption{Traffic Flow Comparison of Nodes \#99, \#100, and \#200.}

  \label{fig_3}
\end{figure}
Beyond computational complexity, another critical challenge is the inherent noise in traffic sensor data. As illustrated in Figure \ref{fig_3}, traffic flow signals exhibit both clear periodic patterns and significant high-frequency fluctuations. The magnitude of this volatility varies considerably across different locations; for instance, the traffic flow of Node \#99 exhibits much greater random fluctuations than Node \#200. These stochastic components, arising from factors like unpredictable incidents, can obscure the underlying trends that are crucial for accurate forecasting. While prior work like SDformer \cite{zhou2024sdformer} has addressed this by applying filtering in the frequency domain, its use of a hard top-K selection may introduce external signal distortion. This motivates us to introduce a more robust filtering mechanism that can effectively isolate the primary trends from high-frequency noise without compromising signal integrity.


To address the dual challenges of computational complexity and signal noise in traffic forecasting, this paper introduces a Butterworth-Enhanced Spatial-Temporal Mamba model, named \textbf{ButterMamba}.
The model synergistically integrates two core components: a Butterworth Spectral Filtering (BSF) module to suppress high-frequency noise and isolate meaningful trends, and a Spatial-Temporal State Mixer (STSM) that leverages the parallel Mamba architecture to efficiently capture both long-range temporal dependencies and complex spatial correlations. By decoupling noise filtering from spatial-temporal modeling, ButterMamba achieves state-of-the-art forecasting accuracy while maintaining linear computational complexity $O(n)$. Experimental results on three public datasets demonstrate its superior performance, effectively overcoming the efficiency-accuracy trade-off present in many existing models. The contributions of our paper are summarized as follows:
\begin{itemize}
\item We propose ButterMamba, a novel framework leveraging State Space Models for efficient and robust traffic flow prediction. Our approach simultaneously addresses the high computational cost of attention mechanisms and performance degradation from noisy sensor data.
\item We design a Butterworth Spectral Filtering module that introduces frequency-domain optimization to enhance signal quality. We further design a Spatio-Temporal State Mixer architecture by extending the Mamba framework with directional state transitions to capture complex spatial-temporal dependencies in linear time.
\item We conduct extensive experiments on three public benchmark datasets. Results demonstrate that ButterMamba achieves state-of-the-art prediction accuracy while significantly reducing training time and memory consumption compared to existing models.
\end{itemize}

\section{Related Work}
\subsection{Traffic Prediction with Deep Learning}
Recently, an increasing number of researchers have employed deep learning techniques to traffic flow prediction tasks. In the early stages, researchers attempted to use Convolutional Neural Networks (CNNs) and Recurrent Neural Networks (RNNs), along with their variants, to separately extract spatial and temporal information from spatiotemporal graphs \cite{STGCN}. However, CNNs are not well-suited for capturing the non-Euclidean structure of graphs. 

With the development of Graph Convolutional Networks (GCNs) and novel models such as Transformers, researchers have begun integrating GCNs with advanced temporal models to extract traffic-related features effectively. Jiang et al. \cite{PDFormer} designed a spatial self-attention module to capture dynamic spatial dependencies. By introducing two graph masking matrices, they highlighted the spatial dependencies from both short-range and long-range perspectives. Meanwhile, Gao et al. \cite{ijcai2024p442} proposed two decoupled masked autoencoders to reconstruct spatiotemporal sequences, aiming to address the spatiotemporal mirage problem. However, they did not take into account that for large-scale networks with numerous nodes, such quadratic complexity computational models face prohibitive computation times. 

Given the ever-changing state of urban traffic, incorporating new data into the model and continuing training are essential. However, high-complexity models make the prediction costs increasingly unacceptable.

\subsection{Sequence Modeling with State Space Models}


Originating from the Kalman filter \cite{kalmanfilter}, State Space Models (SSMs) have recently emerged as a promising paradigm for long-sequence modeling due to their powerful and computationally efficient nature. Modern SSMs are formulated as continuous Linear Time-Invariant (LTI) systems, which can be discretized for use in deep learning.

A key breakthrough in applying SSMs to deep learning was the High-order Polynomial Projection Operators (HiPPO) framework \cite{hippo}, which provided a principled way to initialize the SSM matrices to effectively memorize long histories. Based on this, the Structured State Space for Sequences (S4) \cite{S4} model demonstrated that by structuring these matrices, an SSM could be computed with high efficiency as a recurrent or a convolutional model. This allowed S4 to match or exceed the performance of Transformers on challenging long-range benchmarks while being significantly faster at inference.

The Mamba architecture \cite{mamba} was further advanced by introducing a selection mechanism. This mechanism makes the SSM's parameters input-dependent, allowing the model to dynamically adapt its behavior based on the input. This enables Mamba to selectively focus on relevant information and discard irrelevant parts — a capability previously associated with attention mechanisms. Importantly, Mamba achieves this powerful modeling ability while maintaining linear-time complexity with respect to sequence length, thereby overcoming the performance-efficiency trade-off often seen in Transformer-based models. Its success has led to widespread adoption across diverse domains, including vision \cite{visionmamba} and, more recently, traffic flow forecasting.

In particular, Shao et al. \cite{STmamba} proposed the Spatial-Temporal Mamba (ST-Mamba), the first framework to leverage SSMs for traffic flow prediction without relying on graph structures. Similarly, Li et al. \cite{STGmamba} introduced the Spatial-Temporal Graph Mamba (STG-Mamba), which treats traffic networks as dynamic systems and models their evolution through a novel Spatial-Temporal Selective State Space Module (ST-S3M). These recent advances demonstrate that Mamba-based models are particularly well-suited for spatial-temporal forecasting tasks.

\begin{figure*}[t] 
    \centering
    \includegraphics[width=\textwidth]{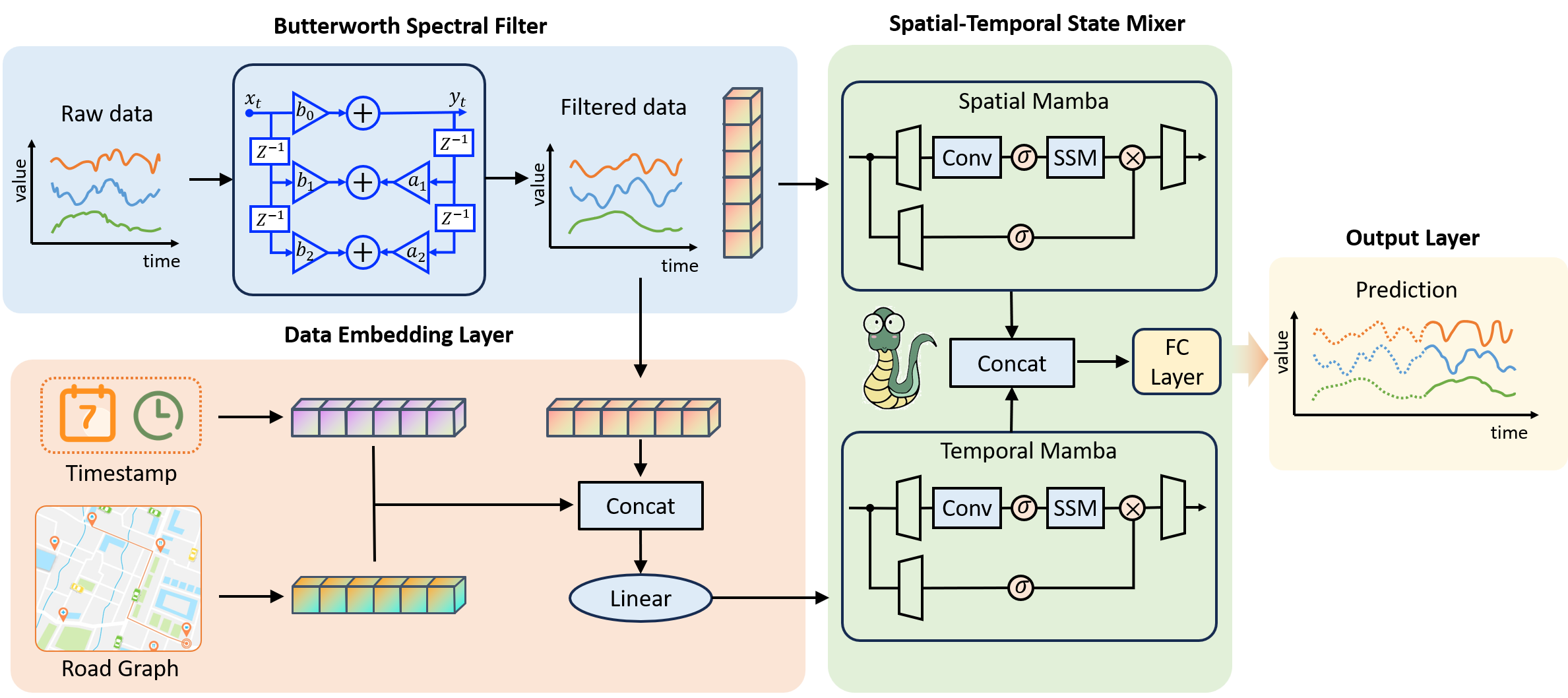} 
    \caption{Overall Framework Diagram of ButterMamba.}
    \label{123}
\end{figure*}

\section{Preliminaries}
\subsection{Road Network Definition}

We model the road network as a graph $\mathcal{G} = (\mathcal{V}, \mathcal{E}, \mathbf{A})$, where $\mathcal{V}$ is a set of $N$ nodes representing traffic sensors, so $|\mathcal{V}| = N$.  $\mathcal{E}$ is a set of edges representing the connections between sensors. $\mathbf{A} \in \mathbb{R}^{N \times N}$ is the weighted adjacency matrix, where $\mathbf{A}_{ij} > 0$ if there is an edge between node $i$ and node $j$, and $\mathbf{A}_{ij} = 0$ otherwise. The edge weights can be derived from the geographical distance between sensors.

\subsection{Problem Definition}

At any given time step $t$, the traffic state on the network is represented by a feature matrix $\mathbf{X}_t \in \mathbb{R}^{N \times C}$, where $N$ is the number of nodes (sensors) and $C$ is the number of traffic features (e.g., speed, flow, occupancy) recorded for each node. The task of traffic flow prediction is to learn a function $f(\cdot)$ that maps a sequence of $T$ historical traffic flow on the graph $\mathcal{G}$ to a sequence of $T'$ future traffic flow. Formally, given the historical observations:
\begin{equation}
\mathcal{X}_{\text{hist}} = (\mathbf{X}_{t-T+1}, \dots, \mathbf{X}_{t})
\end{equation}
The problem is defined as:
\begin{equation}
[\mathcal{X}_{\text{hist}}, \mathcal{G}] \xrightarrow{f} [\mathbf{X}_{t+1}, \dots, \mathbf{X}_{t+T'}]
\end{equation}

\subsection{State Space Models}
A State Space Model (SSM) maps a 1D input signal $x(t)$ to a 1D output signal $y(t)$ through a latent state vector $h(t) \in \mathbb{R}^{D}$. The continuous-time formulation is defined by a pair of linear ordinary differential equations (ODEs):
\begin{align}
    h'(t) &= \mathbf{A}h(t) + \mathbf{B}x(t) \label{eq:ssm_continuous_state} \\
    y(t) &= \mathbf{C}h(t) + \mathbf{D}x(t) \label{eq:ssm_continuous_output}
\end{align}
where $h'(t)$ is the time derivative of the state $h(t)$, $\mathbf{A} \in \mathbb{R}^{D \times D}$ and $\mathbf{B}, \mathbf{C}^\top, \mathbf{D} \in \mathbb{R}^{D \times 1}$ are learnable parameters. 
For simplicity, the skip connection term $\mathbf{D}$ is often omitted in deep learning models.

To be used with discrete sequence data, the continuous system must be discretized. This is achieved by introducing a timescale parameter $\Delta$ and transforming the continuous parameters $(\mathbf{A}, \mathbf{B})$ into discrete parameters $(\overline{\mathbf{A}}, \overline{\mathbf{B}})$. A standard method for this conversion is the Zero-Order Hold (ZOH) \cite{S4}, defined as:
\begin{align}
    \overline{\mathbf{A}} &= \exp(\Delta \mathbf{A}) \label{eq:ssm_discrete_A} \\
    \overline{\mathbf{B}} &= \mathbf{A}^{-1} \left( \exp(\Delta \mathbf{A}) - \mathbf{I} \right) \mathbf{B} \label{eq:ssm_discrete_B}
\end{align}
This yields a discrete recurrence relation for a sequence $x_k$ sampled at time steps $k$:
\begin{align}
    h_k &= \bar{\mathbf{A}}h_{k-1} + \bar{\mathbf{B}}x_k \label{eq:ssm_discrete_state} \\
    y_k &= \mathbf{C}h_k \label{eq:ssm_discrete_output}
\end{align}
This recurrent formulation allows the SSM to be computed efficiently, step-by-step. Alternatively, the full output sequence can be computed in parallel as a single large convolution using a structured convolutional kernel $\overline{\mathbf{K}}$:
\begin{equation}
    y = x * \overline{\mathbf{K}} \label{eq:ssm_convolution}
\end{equation}
where $\overline{\mathbf{K}} \in \mathbb{R}^{L}$ for a sequence of length $L$ is derived from the SSM parameters and is defined as:
\begin{equation}
    \overline{\mathbf{K}} = (\mathbf{C}\overline{\mathbf{B}}, \mathbf{C}\overline{\mathbf{A}}\overline{\mathbf{B}}, \dots, \mathbf{C}\overline{\mathbf{A}}^{L-1}\overline{\mathbf{B}}) \label{eq:ssm_kernel}
\end{equation}
This dual representation as both a recurrent and a convolutional model is a central property of modern SSMs, such as Mamba \cite{mamba}, enabling them to be trained in parallel like a CNN and used for efficient autoregressive inference like an RNN.

\section{Methodology}

The overall architecture is illustrated in Figure \ref{123}. The data flows sequentially through a spectral filtering module, a contextual embedding layer, the core spatial-temporal mixer, and a final output layer.

\subsection{Butterworth Spectral Filtering (BSF)}

\begin{figure}[h]
  \centering
  \begin{minipage}[b]{0.45\textwidth}
    \centering
    \includegraphics[width=1.0\linewidth]{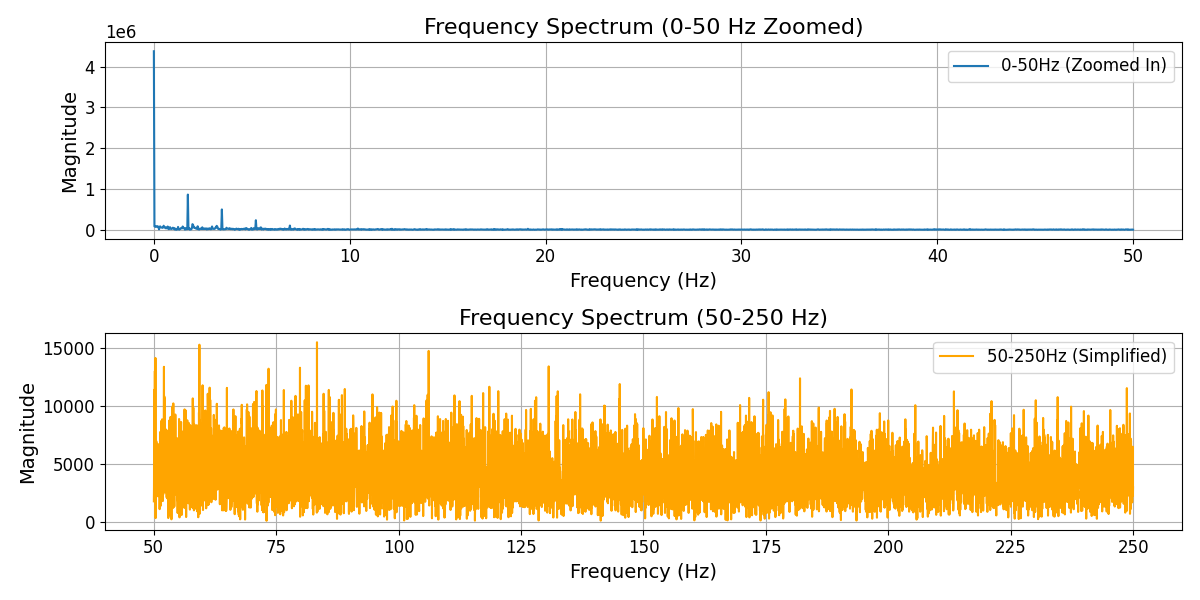}
  \end{minipage}
  
  \caption{Frequency Spectrum of a Raw Traffic Signal.}

  \label{fig_4}
\end{figure}

Raw traffic data is often corrupted by high-frequency fluctuations from sensor errors or stochastic events. As shown by the frequency spectrum analysis in Figure \ref{fig_4}, the vast majority of the signal's power is concentrated in the low-frequency domain, corresponding to the primary periodic and trend components, while the high-frequency domain contains a low-magnitude, noise-like signal. Our BSF module acts as a low-pass filter to mitigate this issue.

We specifically choose the Butterworth filter \cite{selesnick2002generalized} for its maximally flat frequency response in the passband. This property is highly desirable as it ensures that essential low-frequency signals representing daily trends are preserved with minimal amplitude distortion, while high-frequency noise is smoothly attenuated. 

We implement this using a second-order IIR (Infinite Impulse Response) filter, which provides a good balance between effective frequency cutoff and computational efficiency. A higher order could offer a sharper cutoff but at the cost of increased complexity and potential instability, making the second-order design a robust and standard choice. The filtering is applied to each node's time series independently using the following difference equation:

\begin{equation}
    y_t = b_0 x_t + b_1 x_{t-1} + b_2 x_{t-2} - a_1 y_{t-1} - a_2 y_{t-2}
\end{equation}
where $x_t$ is the input signal at time $t$, and $y_t$ is the filtered output. The filter coefficients $(a_i, b_i)$ are determined by the normalized cutoff frequency $f_c \in (0, 1)$, which is the ratio of the desired cutoff frequency to the Nyquist frequency ($f_s/2$, where $f_s$ is the sampling frequency).

The coefficients are calculated via a bilinear transform with pre-warping. First, the normalized digital frequency $f_c$ is pre-warped to its analog equivalent $\Omega_c$:
\begin{equation}
    \Omega_c = \tan\left(\frac{\pi f_c}{2}\right)
\end{equation}
Then, the coefficients are computed as follows:
\begin{equation}
\begin{aligned}
    b_0 &= \frac{\Omega_c^2}{1 + \sqrt{2}\Omega_c + \Omega_c^2} \\
    b_1 &= 2b_0 \\
    b_2 &= b_0 \\
    a_1 &= \frac{2(\Omega_c^2 - 1)}{1 + \sqrt{2}\Omega_c + \Omega_c^2} \\
    a_2 &= -\frac{1 - \sqrt{2}\Omega_c + \Omega_c^2}{1 + \sqrt{2}\Omega_c + \Omega_c^2}
\end{aligned}
\end{equation}

Note that $a_i$ are defined as the negative of the standard Butterworth denominator coefficients to match the additive feedback structure shown in Figure \ref{123}. The output of this module is the denoised feature tensor, $\mathbf{X}_{\text{filtered}}$, which serves as a cleaner, more stable input for the subsequent model layers.

\subsection{Data Embedding Layer}
The Data Embedding Layer enriches the filtered traffic features with essential contextual information about their position in time and space.

\subsubsection{Temporal Embedding}
To capture the strong periodic patterns inherent in traffic flow (e.g., daily and weekly cycles), we generate temporal embeddings from the time of day and the day of the week. For each time step $t$, we obtain its time-of-day index (e.g., 0 to 287 for 5-minute intervals) and its day-of-week index (0 to 6). These two indices are passed through separate embedding layers and then concatenated to form the final temporal feature vector, $\mathbf{T}_{\text{emb}} \in \mathbb{R}^{N \times D_t}$, where $D_t$ is the dimension of the temporal embedding.

\subsubsection{Spatial Embedding}
To encode each node's position and structural role within the road network topology, we employ Laplacian Eigenmaps \cite{belkin2003laplacian}, a powerful technique for learning low-dimensional representations of graph nodes. This process involves the following steps:
\begin{enumerate}
    \item We first compute the symmetric normalized Laplacian matrix of the graph $\mathcal{G}$:
    \begin{equation}
        \mathbf{L}_{\text{norm}} = \mathbf{I} - \mathbf{D}^{-1/2}\mathbf{A}\mathbf{D}^{-1/2}
    \end{equation}
    where $\mathbf{A}$ is the adjacency matrix, $\mathbf{D}$ is the diagonal degree matrix, and $\mathbf{I}$ is the identity matrix.

    \item We then perform an eigendecomposition of the Laplacian matrix: $\mathbf{L}_{\text{norm}} = \mathbf{U}\mathbf{\Lambda}\mathbf{U}^\top$, where $\mathbf{U}$ is the matrix of eigenvectors and $\mathbf{\Lambda}$ is the diagonal matrix of eigenvalues.

    \item 
We select the $k$ eigenvectors corresponding to the smallest non-trivial eigenvalues from $\mathbf{U}$ and use a linear projection to generate the final spatial embedding, $\mathbf{S}_{\text{emb}} \in \mathbb{R}^{N \times D_s}$. In our implementation, we set $k=3$.
\end{enumerate}

\subsubsection{Feature Fusion}
Finally, 
the spatial embedding $\mathbf{S}_{\text{emb}}$ and temporal embedding $\mathbf{T}_{\text{emb}}$ are concatenated with the filtered traffic features, $\mathbf{X}_{\text{filtered}}$, along the feature dimension. This fused tensor is then passed through a final linear layer to project it to the model's hidden dimension, creating the final input representation $\mathbf{X}_{\text{emb}}$ for the subsequent layers:
\begin{equation}
\mathbf{X}_{\text{emb}} = \text{Linear}\left(\text{Concat}(\mathbf{X}_{\text{filtered}}, \mathbf{S}_{\text{emb}}, \mathbf{T}_{\text{emb}})\right)
\end{equation}

\subsection{Spatial-Temporal State Mixer (STSM)}


The parallel architecture of our Spatial-Temporal State Mixer (STSM) is motivated by a fundamental observation in traffic dynamics: spatial and temporal dependencies arise from intrinsically distinct physical mechanisms, necessitating decoupled modeling rather than unified processing.

Spatial dependencies are inherently instantaneous: traffic congestion propagates across road networks within minutes (e.g., upstream congestion rapidly induces downstream queues), reflecting cross-node correlations at the same timestamp. This non-sequential relationship should be modeled via lateral scanning across nodes while fixing the temporal dimension. In contrast, temporal dependencies are strictly sequential: each location exhibits unique periodic patterns (e.g., morning/evening peaks) shaped by local context, representing intra-node evolution over time that requires longitudinal scanning along the temporal axis. Unified architectures (e.g., STG-Mamba) flatten spatio-temporal data into a single sequence, forcing the state space model to perform unnatural "jumps" during state propagation (e.g., from node 
$i$ at time $t$ to node $j$ at $t+1$). This disrupts the continuity of both dependency types, i.e., spatial correlations are fragmented by temporal strides, while temporal evolution is interrupted by node switches, creating representational interference that impedes learning.

The STSM leverages the Mamba architecture. The basic Mamba block combines a discrete convolution with a selective SSM:

\begin{equation} \label{eq:mamba_block}
\begin{aligned}
    \text{Mamba}(x) ={}& \text{SSM}_{\text{selective}}\left( \sigma(\text{Conv}(\text{Linear}(x))) \right) \\
    & \otimes \sigma(\text{Linear}(x))
\end{aligned}
\end{equation}
where $\sigma$ is the SiLU activation function, $\otimes$ denotes element-wise multiplication. The convolutional layer allows the SSM to condition its behavior on local context before propagating information over long distances.

Instead of a single, unified block, our STSM employs a parallel architecture with two independent Mamba blocks. This design allows the model to learn distinct representations for time and space, which are then concatenated to produce the final hidden representation $\mathbf{X}_{\text{hid}}$. The overall architecture is formulated as:

\begin{equation} \label{eq:mamba_block}
\begin{aligned}
\mathbf{X}_{\text{hid}} = \text{Concat}\left( \text{Mamba}^{\text{S}}(\mathbf{X}_{\text{filtered}}^{\top}), \text{Mamba}^{\text{T}}(\mathbf{X}_{\text{emb}}) \right)
\end{aligned}
\end{equation}


\textbf{Temporal Mamba:} 
It takes the fully embedded data $\mathbf{X}_{\text{emb}}$ as input. By using the embedded data, it can leverage rich contextual information—such as the time of day, day of the week, and the node's static spatial position—to learn more nuanced and accurate temporal dynamics. 

\textbf{Spatial Mamba:} 
It uses the denoised but non-embedded data $\mathbf{X}_{\text{filtered}}$ as input. Our experiments revealed that including temporal embeddings in the spatial scan introduced ambiguity, hindering the model's ability to learn clear spatial dependencies. By feeding it only the clean traffic values, the Spatial Mamba can focus on learning the direct relationships between traffic flows at different locations at a given moment, without interference from time-varying features.

Finally, we incorporate two practical refinements to enhance model efficacy: 
(1) \textit{Asymmetric state dimensionality}: We allocate distinct state dimensions for temporal (128D) and spatial (16D) dependencies, reflecting their different representational complexity. 
(2) \textit{Input modality separation}: The Spatial Mamba operates solely on denoised traffic values ($X_{\text{filtered}}^T$) without temporal embeddings, while the Temporal Mamba utilizes full contextual embeddings ($X_{\text{emb}}$) to preserve periodic patterns. 
Ablation studies confirm that these design choices collectively contribute to improved prediction accuracy.

\subsection{OutPut layer}
The final stage of the ButterMamba framework is a simple yet effective output Layer that transforms the high-dimensional latent representation from the STSM module into the final multi-step forecast. This layer consists of a single linear projection that directly maps the learned features to the desired output shape:
\begin{equation}
    \hat{\mathbf{X}} = \text{Linear}(\mathbf{X}_{\text{hid}})
\end{equation}
where $\mathbf{X}_{\text{hid}}$ is the concatenated hidden representation from the STSM module, and the linear layer is configured to produce the final prediction tensor $\hat{X}\in\mathbf{R}^{N\times C \times L}$ . Here, $N$ is the number of nodes (sensors), $C$ is the number of traffic features, and $L$ is the prediction horizon.

\section{Experiments}
\begin{table}
  \caption{Data description}
  \label{tab:freq}
  \begin{tabular}{cccc}
    \toprule
    Datasets&\#Nodes&\#Interval&Time range\\
    \midrule
    PeMS04&307&5min&01/01/2018-02/28/2018\\
    PeMS07&883&5min&05/01/2017-08/31/2017\\
    PeMS08&170&5min&07/01/2016-08/31-2016\\
  \bottomrule
\end{tabular}
\end{table}

\subsection{Datasets}
To evaluate the performance of our work, we conduct experiments on three public real-world traffic datasets, i.e., PeMS04, PeMS07, PeMS08 \cite{song2020spatial}. PeMS means Caltrans Performance Measure System (PeMS) \cite{chen2001freeway} , and other details are given in Table \ref{tab:freq}.
\subsection{Baselines}
We compare our model with the following baselines methods.They are \textbf{ARIMA} \cite{fang2021spatial}, \textbf{VAR} \cite{song2020spatial}, \textbf{SVR} \cite{song2020spatial}, \textbf{LSTM} \cite{song2020spatial}, \textbf{TCN} \cite{lan2022dstagnn}, \textbf{Transformer} \cite{vaswani2017attention}, \textbf{DCRNN} \cite{li2017diffusion}, \textbf{STGCN} \cite{STGCN}, \textbf{ASTGCN} \cite{guo2019attention}, 
\textbf{GWNet} \cite{wu2019graph}, 
\textbf{STGODE} \cite{fang2021spatial}, 
\textbf{DSTAGNN} \cite{lan2022dstagnn}, 
\textbf{ST-WA} \cite{cirstea2022towards}, 
\textbf{ASTGNN} \cite{guo2021learning}, 
\textbf{AGCRN} \cite{bai2020adaptive}, 
\textbf{STNorm} \cite{deng2021st}, 
\textbf{PDFormer} \cite{PDFormer}, 
\textbf{STAEformer} \cite{liu2023spatio},
\textbf{STD-MAE} \cite{gao2023spatial},
\textbf{ST-Mamba} \cite{STmamba}, 
\textbf{DTRformer} \cite{Chen_2025}, \textbf{STG-Mamba} \cite{STGmamba}.
\subsection{Experimental settings}

\textbf{Dataset Processing.} 
To ensure a fair comparison with prior work, we adopt a standardized data processing and evaluation protocol for all datasets. Each dataset is split into training, validation, and test sets using a 6:2:2 ratio. To prevent data leakage in the validation and test sets, we apply filtering operations separately to each of the three subsets. We formulate the forecasting task as a sequence-to-sequence problem, where the model uses one hour of historical data (12 time steps at 5-minute intervals) to predict the subsequent hour (12 future time steps). Before training, the traffic data in the training set is normalized using Min-Max scaling. 


\textbf{Model Settings.} 
All experiments are conducted on a machine equipped with a single NVIDIA GeForce RTX 4090 GPU and 256 GB of RAM. The operating environment is Ubuntu 18.04 with CUDA 11.8, PyTorch 2.0.0, and Python 3.10. The state dimensions of the temporal-Mamba and spatial-Mamba SSMs are set to 128 and 16, respectively. The number of linear layers in the final decoding stage is searched over {108, 72, 36, 1}. Additionally, to prevent overfitting, we apply a dropout \cite{srivastava2014dropout} function with a rate of 0.2. We train our model using AdamW optimizer \cite{loshchilov2017fixing} with a learning rate of 0.001. The batch size is 512, and the training epoch is 250. The optimal model is determined based on the performance in the validation set.

\textbf{Evaluation Metrics.}
As with other traffic flow prediction models, we adopt three evaluation metrics Mean Absolute Error (MAE), Mean Absolute Percentage Error (MAPE), and Root Mean Squared Error (RMSE) to quantify the prediction accuracy across all compared methods.

\begin{table*}[ht]
\centering
\caption{Model performance comparison on different datasets.}
\label{tab:results}
\small
\begin{tabular}{l|lll|lll|lll}
\toprule
\multirow{2}{*}{Model}  & \multicolumn{3}{c|}{PeMS04} & \multicolumn{3}{c|}{PeMS07} & \multicolumn{3}{c}{PeMS08} \\
\cmidrule(lr){2-4} \cmidrule(lr){5-7} \cmidrule(lr){8-10} \\
& MAE & RMSE & MAPE & MAE & RMSE & MAPE & MAE & RMSE & MAPE \\
\midrule
\textbf{ARIMA} \cite{fang2021spatial}& 33.73 & 48.80 & 24.18 & 38.17 & 59.27 & 19.46 & 31.09 & 44.32 & 22.73 \\
\textbf{VAR} \cite{song2020spatial} & 23.75 & 36.66 & 18.09 & 75.63 & 115.24 & 32.22 & 23.46 & 36.33 & 15.42 \\
\textbf{SVR} \cite{song2020spatial} & 28.70 & 44.56 & 19.20 & 32.49 & 50.22 & 14.26 & 23.25 & 36.16 & 14.64 \\
\textbf{LSTM} \cite{song2020spatial}  & 27.14 & 41.59 & 18.20 & 29.98 & 45.84 & 13.20 & 22.20 & 34.06 & 14.20 \\
\textbf{TCN} \cite{lan2022dstagnn} & 31.11 & 37.25 & 15.48 & 32.68 & 42.23 & 14.22 & 22.69 & 35.79 & 14.04 \\
\textbf{Transformer} \cite{vaswani2017attention}  & 23.83 & 37.19 & 15.57 & 26.80 & 42.95 & 12.11 & 18.52 & 28.68 & 13.66 \\
\textbf{DCRNN} \cite{li2017diffusion} & 24.70 & 38.12 & 17.12 & 25.30 & 38.58 & 11.66 & 17.86 & 27.83 & 11.45 \\
\textbf{STGCN} \cite{STGCN} & 22.70 & 35.55 & 14.59 & 25.38 & 38.78 & 11.08 & 18.02 & 27.83 & 11.40\\
\textbf{ASTGCN} \cite{guo2019attention} & 22.93 & 35.22 & 16.56 & 28.05 & 42.57 & 13.92 & 18.61 & 28.16 & 13.08\\
\textbf{GWNet} \cite{wu2019graph} & 25.45 & 39.70 & 17.29 & 26.85 & 42.78 & 12.12 & 19.13 & 31.05 & 12.68\\
\textbf{STGODE} \cite{fang2021spatial} & 20.84 & 32.82 & 13.77 & 22.99 & 37.54 & 10.14 & 16.81 & 25.97 & 10.62\\
\textbf{DSTAGNN} \cite{lan2022dstagnn} & 19.30 & 31.46 & 12.70 & 21.42 & 34.51 & 9.01 & 15.67 & 24.77 & 9.94\\
\textbf{ST-WA} \cite{cirstea2022towards} & 19.06 & 31.02 & 12.52 & 20.74 & 34.05 & 8.77 & 15.41 & 24.62 & 9.94\\
\textbf{ASTGNN} \cite{guo2021learning} & 19.26 & 31.16 & 12.65 & 22.23 & 35.95 & 9.25 & 15.98 & 25.67 & 9.97\\
\textbf{AGCRN} \cite{bai2020adaptive} & 19.83 & 32.26 & 12.97 & 21.29 & 35.12 & 8.97 & 15.95 & 25.22 & 10.09\\
\textbf{STNorm} \cite{deng2021st} & 19.21 & 32.30 & 13.05 & 20.59 & 34.86 & 8.61 & 15.39 & 24.80 & 9.91\\
\textbf{PDFormer} \cite{PDFormer} & 18.32 & 29.97 & 12.10 & 19.83 & 32.87 & 8.53 & 13.58 & 23.51 & 9.05\\
\textbf{STAEformer} \cite{liu2023spatio} & 18.22 & 30.18 & 11.98 & 19.14 & 32.60 & 8.01 & 13.46 & 23.25 & 8.88\\
\textbf{STD-MAE} \cite{gao2023spatial} & 17.80 & 29.25 & 11.97 & 18.65 & 31.44 & 7.84 & 13.44 & 22.47 & 8.76\\
\textbf{ST-Mamba} \cite{STmamba} & 18.19 & 30.17 & 11.88 & 19.07 & 32.40 & 8.02 & 13.40 & 23.20 & 9.00 \\
\textbf{DTRformer} \cite{Chen_2025} & 18.00 &  29.58 & 12.30 & 18.99 & 32.23 & 7.93 & \textbf{13.17} & 22.85 & 8.66 \\
\textbf{STG-Mamba} \cite{STGmamba} & 18.09 & 29.53 & 12.11 & - - - & - - - & - - - & - - - & - - - & - - - \\
\midrule
\textbf{Ours} & \textbf{16.71 $\pm$ 0.18} & \textbf{26.72 $\pm$ 0.31} & \textbf{11.21 $\pm$ 0.22} & \textbf{17.94 $\pm$ 0.20} & \textbf{28.55 $\pm$ 0.34} & \textbf{7.59 $\pm$ 0.08} & 13.32 $\pm$ 0.11 & \textbf{20.83 $\pm$ 0.15} & \textbf{8.46 $\pm$ 0.06}\\
\bottomrule
\end{tabular}
\end{table*}

\subsection{Performance Comparison}
The comprehensive multi-step forecasting results are presented in Table \ref{tab:results}. To ensure a fair comparison, the results for all baseline models are cited directly from their original publications or established benchmark papers. To rule out the possibility of random chance and to demonstrate the reproducibility of our results, we conducted 5 independent runs for all reported models using different random seeds. We report the mean $\pm$ standard deviation in Tables. The consistently small standard deviations (e.g., $\pm 0.18$ MAE for ButterMamba on PeMSD4) confirm the stability and robustness of our method across runs. The results clearly demonstrate the superior performance of ButterMamba across all three datasets and all three evaluation metrics.
The results clearly demonstrate the superior performance of ButterMamba across all three datasets and all three evaluation metrics. 

First, ButterMamba significantly surpasses traditional statistical methods like ARIMA and VAR, as well as early deep learning models such as LSTM and TCN. 
This highlights the critical importance of explicitly modeling spatio-temporal dependencies, a capability that these earlier, non-graph-aware models lack. 

When compared to models that utilize graph convolutions, such as STGCN, DCRNN, and GWNet, ButterMamba still shows a clear advantage. For example, ButterMamba achieves an MAE of 17.94 on the PEMS07 dataset, outperforming GWNet (26.85) by a significant margin. This suggests that our STSM module, which uses a sequential scan over nodes via Mamba, is a more effective mechanism for capturing complex spatial correlations than standard graph convolution or diffusion operations.

Meanwhile, ButterMamba shows superior performance over Transformer-based architectures and recent Mamba-based work. Compared to PDFormer and DTRformer on the PEMS04 dataset, our model reduces the MAE by 8.7\% and 7.1\%. 
We also compare ButterMamba against ST-Mamba and STG-Mamba, two recent Mamba-based spatio-temporal models. Though the results of STG-Mamba on PeMSD7 and PeMSD8 are unavailable in the original paper \cite{STGmamba}, our model achieves 16.71(MAE), 26.72 (RMSE), and 11.21 (MAPE) on PeMSD4, outperforming STG-Mamba (18.19, 30.17, 11.88) and ST-Mamba (17.80, 29.25, 11.97) across all three metrics. This demonstrates that, even under fair comparison on a shared benchmark, ButterMamba exhibits consistent improvements.

\begin{table}[h]
\caption{Computational cost comparison on PeMS04.}
\centering
\begin{tabular}{lccc}
\toprule
\multirow{2}{*}{Model} & \multicolumn{2}{c}{Computation time} & GPU Cost \\
\cmidrule(lr){2-3} \cmidrule(lr){4-4}
& epoch & Training(s/epoch) & Memory(MB) \\
\midrule
PDFormer   & 200 & 81.1 & 6286 \\
AGCRN    & 100 & 10.2 & 4167 \\
DSTAGNN   & 107 & 65.1 & 5845 \\
STG-Mamba & 200 & 5.1 & 2749 \\
Ours & 250 & 1.3 & 1260 \\
\bottomrule
\end{tabular}
\end{table}

\subsection{Efficiency Study}
To empirically validate the efficiency of the proposed model, we conduct a direct comparison of the computational cost of ButterMamba against PDFormer, AGCRN, and DSTAGNN. All experiments were conducted on the same server equipped with an AMD Ryzen 3945WX processor and a single NVIDIA RTX 4090 GPU, using consistent batch sizes across experiments.

The results are shown in Table 3. ButterMamba's training time is approximately 1.3 seconds per epoch, which is over 60 times faster than PDFormer, 50 times faster than DSTAGNN, and 8 times faster than AGCRN. Furthermore, our model's peak GPU memory usage is only 1260 MB, a reduction of nearly 80\% compared to PDFormer's 6286 MB. Moreover, we also compare our approach with STG-Mamba, another model based on the Mamba architecture. Our experiments demonstrate that the Mamba architecture is highly competitive in terms of efficiency. By employing a shallower stack with fewer Mamba blocks, our model consumes only half the GPU memory of STG-Mamba and requires approximately one-quarter of the training time per epoch. 

While ButterMamba uses 250 epochs versus 200 for STG-Mamba, optimal training duration is inherently architecture-dependent and determined by validation-based early stopping (not arbitrary epoch counts). Importantly, total training time (not epoch number) is the practical efficiency metric. Even with 25\% more epochs, ButterMamba's total training time remains 3.1× lower (325s vs. 1,020s) with 54\% less GPU memory consumption. This advantage stems from our linear-complexity SSM scans and decoupled processing, not extended training. Per-epoch learning efficiency (accuracy gain per second) thus favors ButterMamba, making it more suitable for real-world deployment where wall-clock time matters most.


\subsection{Ablation Study}
To validate the effectiveness of the individual components within our ButterMamba framework, we conduct an ablation study on the PeMS04 dataset. We compare the full model against three variants, each with a key component removed:
\begin{itemize}
    \item \textbf{w/o Filter:} We remove the BSF module and feed the raw, normalized traffic data directly into the model. 
    \item \textbf{w/o Spatial-Mamba (S-M):} We remove the Spatial Mamba block from the STSM module. 
    \item \textbf{w/o Temporal-Mamba (T-M):} We remove the Temporal Mamba block from the STSM module. 
\end{itemize}

\begin{figure}[h]
  \centering
  \begin{minipage}[b]{0.5\textwidth}
    \centering
    \includegraphics[width=1.0\linewidth]{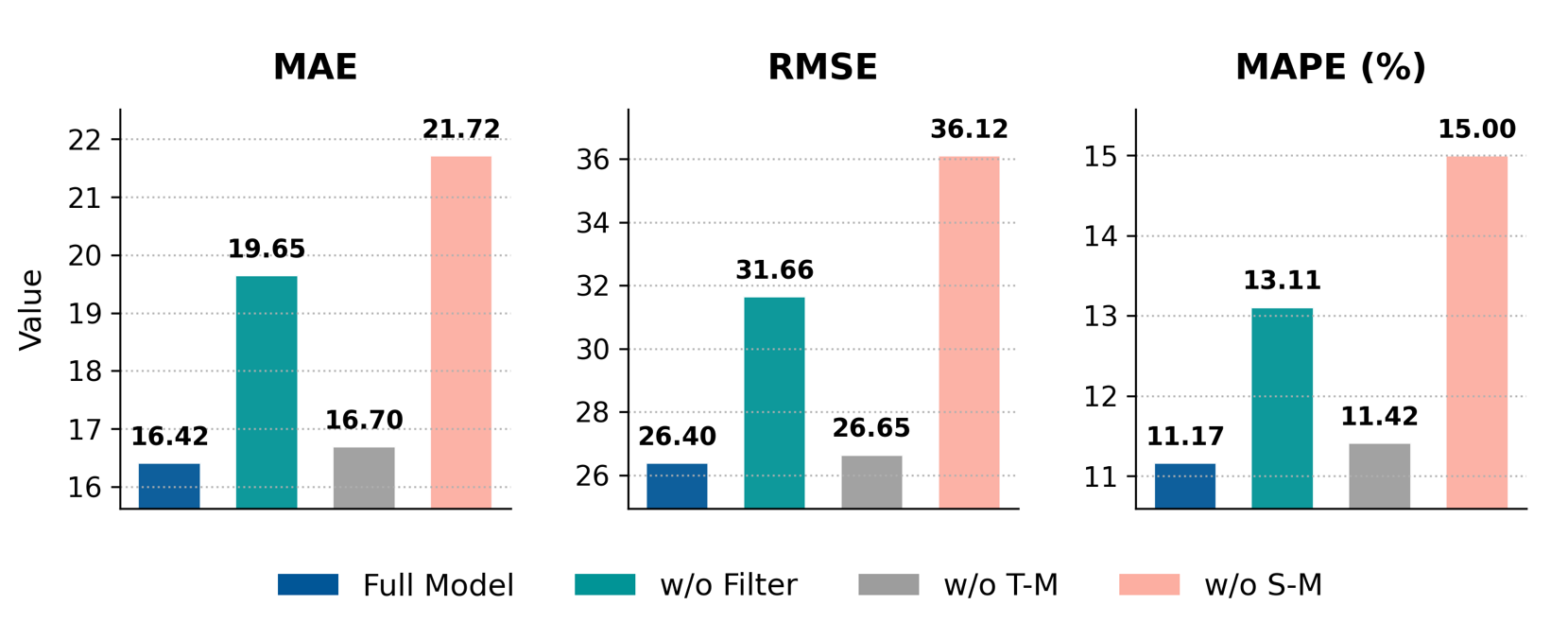}
  \end{minipage}
  
  \caption{Ablation Study on PeMS04. }

  \label{fig_45}
\end{figure}

The results of this study are presented in Figure \ref{fig_45}. From these results, we draw the following key conclusions:

\textbf{The BSF module is crucial for performance.} Removing the BSF module causes a severe degradation in performance across all metrics. 
This strongly validates our hypothesis that filtering high-frequency noise is essential for allowing the model to learn robust and generalizable trends.

\textbf{Spatial dependencies are the dominant factor in prediction.} The removal of the spatial modeling performs the worst. In contrast, the removal of the temporal block sees only a minor increase in MAE. This suggests that for predicting a node's future state, the concurrent traffic conditions across the network provide significantly more predictive power than its own isolated history, aligning with the real-world nature of traffic propagation.

\textbf{Both STSM components are beneficial.} While the spatial component is clearly dominant, the full ButterMamba model still outperforms the other variants. This demonstrates that the dedicated temporal Mamba block, while less critical than the spatial one, captures complementary information that refines the final prediction. The synergistic combination of both modules in our parallel STSM architecture is what leads to optimal performance.

\subsection{Parameter Sensitivity Study}

Given that the cutoff frequency is a critical hyperparameter (as too low a value risks discarding informative signal components, whereas too high a value risks retaining undesirable noise), we conduct a dedicated sensitivity analysis to investigate its impact. As the traffic data in our setting lacks a physically defined sampling rate (and thus an absolute cutoff frequency in Hz), we adopt a practical normalization strategy: assuming a nominal sampling frequency of 500 Hz, we vary the normalized cutoff frequency to simulate different filtering strengths. This allows us to systematically evaluate model performance under varying degrees of high-frequency suppression. 

\begin{figure}[h]
  \centering
  \begin{minipage}[b]{0.5\textwidth}
    \centering
    \includegraphics[width=1.0\linewidth]{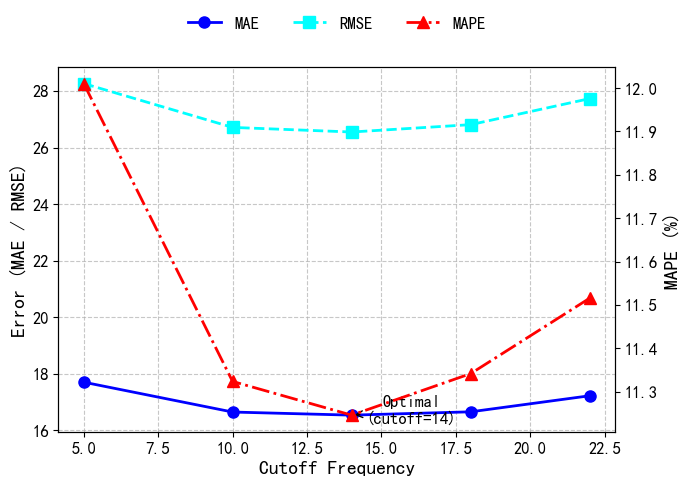}
  \end{minipage}
  
  \caption{Parameter Sensitivity Study on PeMS04.}

  \label{sens}
\end{figure}

As illustrated in Figure \ref{sens}, the model exhibits a clear performance trend: prediction errors (MAE, RMSE, and MAPE) first decrease and then increase as the cutoff frequency varies from 5 to 22 Hz. The optimal performance is achieved at a cutoff period of 14 Hz, yielding an MAE of 16.54, RMSE of 26.55, and MAPE of 11.25\%. Notably, the performance remains stable within a broad range (10–18 Hz), with all metrics fluctuating by less than 0.3. This demonstrates that ButterMamba is not only effective but also robust to reasonable choices of the cutoff frequency, which aligns with the typical short-term periodicity observed in urban traffic flow patterns.
\subsection{Case Study}
\begin{figure}[h]
  \centering
  \begin{minipage}[b]{0.48\textwidth}
    \centering
    \includegraphics[width=1.0\linewidth]{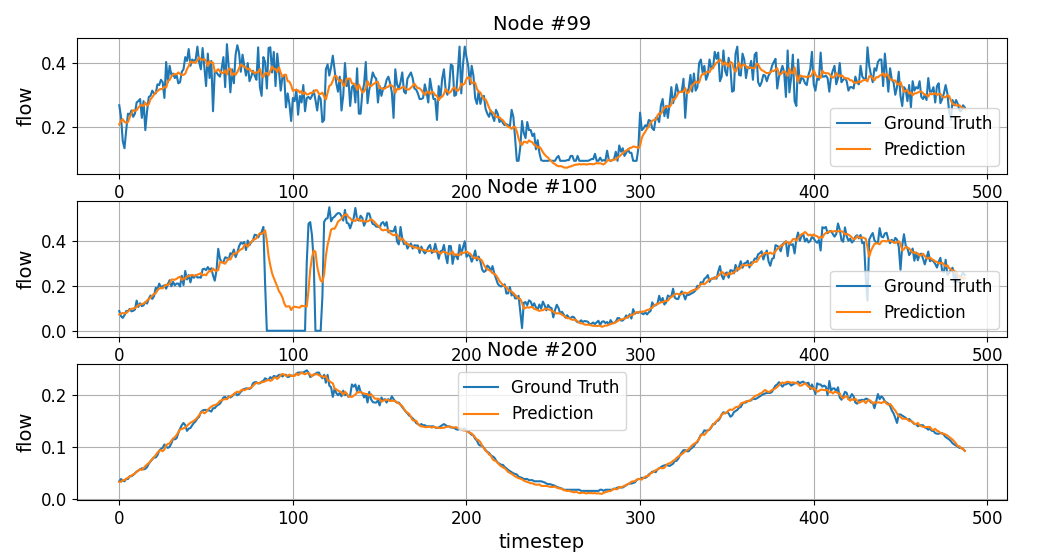}
  \end{minipage}
  
  \caption{Visualization of Traffic Prediction on PeMS04. }

  \label{fig_5}
\end{figure}

To gain a comprehensive understanding of ButterMamba's behavior, we visualize its prediction results against the ground truth for three representative nodes, each exhibiting a different traffic pattern, as illustrated in Figure \ref{fig_5}. 

The adaptive noise suppression mechanism effectively mitigates high-frequency fluctuations in various road segments. Node \#99 displays highly volatile traffic flow, with constant, noisy fluctuations. The plot shows that our model's prediction (orange line) successfully captures the primary diurnal trend without overfitting to the high-frequency noise. Node \#100 presents a challenging scenario with sudden, sharp drops and spikes in traffic flow (e.g., around timestep 80). While the model does not perfectly replicate the magnitude of these extreme, near-instantaneous changes, it correctly identifies their occurrence and quickly recovers to track the subsequent trend. This indicates that the selective mechanism of the Mamba architecture is robust enough not to be ``derailed'' by transient, unpredictable events, instead maintaining its focus on the overall traffic state. Node \#200 shows a smooth, clean, and highly periodic traffic pattern. For this stable signal, ButterMamba's predictions are remarkably accurate, closely tracking the ground truth throughout the morning peak and evening trough. 


These observations demonstrate that ButterMamba prioritizes macroscopic regularities over microscopic perturbations, making it particularly suitable for traffic systems where periodicity and trend dominance prevail over random fluctuations.

\section{Conclusion}

In this work, we introduce ButterMamba, designed to overcome the critical challenges of computational cost and signal noise in traffic forecasting. By synergistically integrating a Butterworth Spectral Filtering (BSF) module for denoising and a parallel Spatial-Temporal State Mixer (STSM) built on the efficient Mamba architecture, our model achieves excellent performance in terms of both accuracy and efficiency. Experiments on three public benchmarks confirm that ButterMamba significantly outperforms state-of-the-art baselines, while significantly reducing training time and memory usage. This superior performance validates our design, showing the benefits of decoupling signal pre-processing from the learning of spatial-temporal dependencies.

The high efficiency and accuracy of ButterMamba make it a practical solution for deployment in large-scale, real-time intelligent transportation systems. However, the model currently relies on a static graph structure and a fixed filtering frequency. Promising directions include integrating dynamic graph learning mechanisms and developing adaptive filtering techniques to further enhance the model's robustness and applicability to a broader range of spatial-temporal forecasting problems.

\bibliographystyle{ACM-Reference-Format}
\bibliography{sample-base}


\end{document}